\documentclass[preprint]{elsarticle}
\usepackage[]{graphicx}
\usepackage[]{xcolor}
\makeatletter
\def\maxwidth{ %
  \ifdim\Gin@nat@width>\linewidth
    \linewidth
  \else
    \Gin@nat@width
  \fi
}
\makeatother

\definecolor{fgcolor}{rgb}{0.345, 0.345, 0.345}
\newcommand{\hlnum}[1]{\textcolor[rgb]{0.686,0.059,0.569}{#1}}%
\newcommand{\hlstr}[1]{\textcolor[rgb]{0.192,0.494,0.8}{#1}}%
\newcommand{\hlcom}[1]{\textcolor[rgb]{0.678,0.584,0.686}{\textit{#1}}}%
\newcommand{\hlopt}[1]{\textcolor[rgb]{0,0,0}{#1}}%
\newcommand{\hlstd}[1]{\textcolor[rgb]{0.345,0.345,0.345}{#1}}%
\newcommand{\hlkwb}[1]{\textcolor[rgb]{0.69,0.353,0.396}{#1}}%
\newcommand{\hlkwc}[1]{\textcolor[rgb]{0.333,0.667,0.333}{#1}}%
\newcommand{\hlkwd}[1]{\textcolor[rgb]{0.737,0.353,0.396}{\textbf{#1}}}%

\usepackage{framed}
\makeatletter
\newenvironment{kframe}{%
 \def\at@end@of@kframe{}%
 \ifinner\ifhmode%
  \def\at@end@of@kframe{\end{minipage}}%
  \begin{minipage}{\columnwidth}%
 \fi\fi%
 \def\FrameCommand##1{\hskip\@totalleftmargin \hskip-\fboxsep
 \colorbox{shadecolor}{##1}\hskip-\fboxsep
     \hskip-\linewidth \hskip-\@totalleftmargin \hskip\columnwidth}%
 \MakeFramed {\advance\hsize-\width
   \@totalleftmargin\z@ \linewidth\hsize
   \@setminipage}}%
 {\par\unskip\endMakeFramed%
 \at@end@of@kframe}
\makeatother

\definecolor{shadecolor}{rgb}{.97, .97, .97}
\definecolor{messagecolor}{rgb}{0, 0, 0}
\definecolor{warningcolor}{rgb}{1, 0, 1}
\definecolor{errorcolor}{rgb}{1, 0, 0}
\newenvironment{knitrout}{}{} 

\usepackage{alltt}

\usepackage{listings}
\usepackage{lineno}
\usepackage{hyperref}
\usepackage{booktabs}
\usepackage{siunitx}
\usepackage{graphicx}
\usepackage{xcolor}
\usepackage{enumitem}
\usepackage{csquotes}
\usepackage{amsmath}

\graphicspath{ {./} }

\journal{Environmental Modelling and Software}

\biboptions{authoryear}

\IfFileExists{upquote.sty}{\usepackage{upquote}}{}
\begin{document}

\begin{frontmatter}

\title{rFIA: An R package for estimation of forest attributes with the US Forest Inventory and Analysis Database}

\author[1]{Hunter Stanke\corref{cor1}}
\ead{stankehu@msu.edu}

\author[2]{Andrew O. Finley}
\author[3]{Aaron S. Weed}
\author[4]{Brian F. Walters}
\author[4]{Grant M. Domke}

\cortext[cor1]{Corresponding author}

\address[1]{Department of Forestry, Michigan State University, East Lansing, MI, USA}
\address[2]{Department of Geography, Michigan State University, East Lansing, MI, USA}
\address[3]{Northeast Temperate Inventory and Monitoring Network, National Park Service, Woodstock, VT, USA}
\address[4]{Forest Service, Northern Research Station, US Department of Agriculture, 1992 Folwell Avenue, St Paul, MN 55108, US}

\begin{abstract}
Forest Inventory and Analysis (FIA) is a US Department of Agriculture Forest Service program that aims to monitor changes in forests across the US. FIA hosts one of the largest ecological datasets in the world, though its complexity limits access for many potential users. \texttt{rFIA} is an R package designed to simplify the estimation of forest attributes using data collected by the FIA Program. Specifically, \texttt{rFIA} improves access to the spatio-temporal estimation capacity of the FIA Database via space-time indexed summaries of forest variables within user-defined population boundaries (e.g., geographic, temporal, biophysical). The package implements multiple design-based estimators, and has been validated against official estimates and sampling errors produced by the FIA Program. We demonstrate the utility of \texttt{rFIA} by assessing changes in abundance and mortality rates of ash (\emph{Fraxinus spp.}) populations in the Lower Peninsula of Michigan following the establishment of emerald ash borer (\emph{Agrilus planipennis}). 
\end{abstract}

\begin{keyword}
\texttt{Forest Inventory and Analysis, R package, \texttt{rFIA}}
\end{keyword}

\end{frontmatter}


\section{Introduction}
Forest Inventory and Analysis (FIA) is a United States Department of Agriculture Forest Service program with the goal of monitoring and projecting changes in forests across the United States (US) \citep{fiaMain}. The program collects, publishes, and analyzes data describing the extent, condition, volume, growth, and use of trees from all land ownerships in the nation, with some records dating back to the early 1930s \citep{tinkham2018applications, smith2002forest}. In 1999, the FIA program established a systematic grid of permanent ground plots that are evenly divided into panels measured in a continuous cycle, allowing spatially unbiased estimates of forest attributes to be computed on an annual basis \citep{gillespie1999rationale, smith2002forest}. The flexibility inherent to the FIA inventory system creates unprecedented opportunities to assess forest change across space, through time, and within unique populations of interest \citep{tinkham2018applications, gray2012forest}.

The research significance of the FIA program has increased in recent decades, with primary applications in carbon cycling, forest growth, forest health, and remote sensing \citep{tinkham2018applications}. The FIA program hosts one of the largest ecological datasets in the world by spatial and temporal extent \citep{tinkham2018applications}, including records from over 5.8 million trees measured at least twice and encompassing a range of ecological diversity unmatched by any other large-scale national forest inventory system \citep{tomppo2010national}. The breadth of attributes currently measured by the FIA program, ranging from forest composition and structure to soil chemistry and invasive species abundance, makes it a vast and powerful resource for monitoring the status and trends in forest attributes from the scale of individual trees to subcontinents \citep{tinkham2018applications}. 


The primary limitation for individuals using FIA data in their own analyses is the complex sample design, database structure, and Structured Query Language used by the FIA program \citep{tinkham2018applications, kromroy2008using}. FIA data are publicly available in several formats (e.g., FIA DataMart) and estimation is facilitated through online tools (e.g., EVALIDator) \citep{toolsData, popsUserGuide} and a non-public R package, FIESTA \citep{fiesta2015}. That said, FIA data are difficult for many non-FIA users to interpret and understand \citep{tinkham2018applications} and there is potential to increase the use of these incredibly rich data among industry professionals, academic scientists, and the general public. To promote use of this valuable public resource and extend the reach of the FIA program and the publicly available data, there is a need for a flexible, user-friendly tool that simplifies the process of working with FIA data for experts and novices alike. To this end, we developed \texttt{rFIA}, an add-on package for R \citep{r2018}.

\texttt{rFIA} implements FIA's design-based estimation procedures \citep{bechtold2005enhanced} for over 50 forest attributes using a simple, yet powerful design. \texttt{rFIA} greatly improves access to the spatio-temporal estimation capacity inherent to the FIA program by allowing space-time indexed summaries of forest attributes to be produced within user-defined population boundaries (e.g., geographic, temporal, biophysical). The package enhances the value of FIA for temporal change detection and forest health monitoring by implementing five design-based estimators that offer flexibility in a balance between precision and temporal smoothing. Our intention in developing \texttt{rFIA} is to provide a versatile, user-friendly software that allows all R users to unlock the value of the FIA program.

\section{Methods}
\subsection{Software design}
We designed \texttt{rFIA} to be intuitive to use and support common data representations by directly integrating other popular R packages into our development. We achieve efficient joins, queries, and data summaries with the \texttt{dplyr} package \citep{wickham2015dplyr}. Specifically, we leverage \texttt{dplyr} for joining and filtering FIA tables, facilitating hierarchical grouping of summary attributes, and implementing non-standard evaluation in \texttt{rFIA} core functions. We achieve efficient space-time query and summary within user-defined population boundaries (i.e., spatial polygons) with the \texttt{sf} package \citep{sf2018}. Parallel processing is implemented with the \texttt{parallel} package \citep{parallel2018}. Parallel implementation is achieved using a snow type cluster \citep{snow} on any Windows OS, and with multi-core forking \citep{parallel2018} on any Unix or Mac OS.

\begin{table}[t!]
  \begin{center}
  \textbf{Software Availability}
    \begin{tabular}{|l|} 
    \hline
\emph{Name of software}  \href{http://rfia.netlify.com/}{rFIA} \\
\emph{Type of software}  Add-on package for R \\
\emph{First available}  2019 \\
\emph{Program languages}  R \\
\emph{License}  GPL 3 \\
\emph{Code Repository} \url{https://cran.r-project.org/web/packages/rFIA/index.html}  \\
\emph{Installation in R} \texttt{install.packages("rFIA")} \\
\emph{Developers}  Hunter Stanke, Andrew O. Finley \\
\emph{Contact Address}  Department of Forestry, \\ \hspace{2.6cm} Michigan State University, East Lansing, MI, USA \\
       \hline
    \end{tabular}
  \end{center}
\end{table}

\subsection{Sampling and estimation procedures}

The design-based estimation procedures used by the FIA program and implemented by \texttt{rFIA} have been widely described in the literature \citep{bechtold2005enhanced, mcroberts2005enhanced, hoffman2014survey}, and hence will only be briefly described here. All estimators for population totals, ratios, and associated variances are derived from the  Horvitz-Thompson estimator and hence incorporate design information via inverse-probability weighting \citep{horvitz1952}. \cite{bechtold2005enhanced} describes in detail the theoretical basis for estimators used by the FIA program and implemented by \texttt{rFIA}.

The FIA program conducts forest inventories using a multi-phased sampling procedure designed to reduce variance through stratification \citep{bechtold2005enhanced, mcroberts2005enhanced}. In the pre-field phase, remotely sensed imagery are used to stratify land area by determining dominant land use in each pixel within the population of interest. In the core phase, permanent ground plots are systematically distributed across the US at a rate of approximately 1 plot per 6000 acres using a hexagonal sampling frame \citep{bechtold2005enhanced, mcroberts2005enhanced}. Each plot is assigned to a single stratum based on the the pre-field stratification of the plot center. If any portion of a plot is determined to contain a forest land use from pre-field stratification, a field crew will visit the site and measure core FIA variables \citep{bechtold2005enhanced}. In the intensive phase, additional forest and ecosystem health variables are measured on 5-15\% of established core plots (approximately 1 plot per 96,000 acres). 

\begin{figure}[t]
    \centering
    \caption{FIA ground plot design (left) and an example map of forested condition classes on a subplot (right).}
    \label{fig:plotDesign}
    \includegraphics[width=14cm]{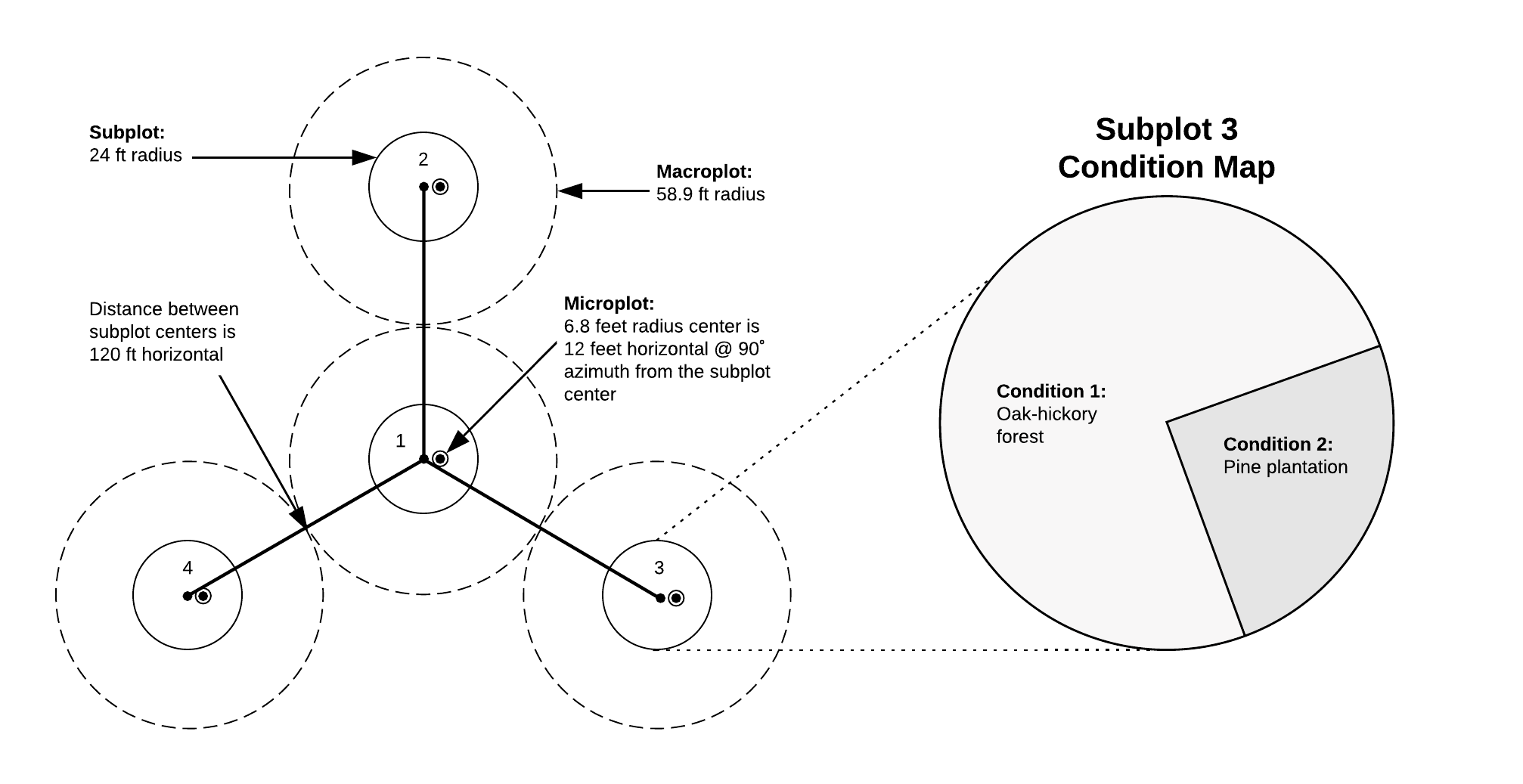}
\end{figure}

FIA permanent ground plots consist of clusters of four subplots (Figure \ref{fig:plotDesign}), where tree attributes are measured for all stems 5.0 inches diameter at breast height (DBH) and larger. Within each subplot, a microplot is established where tree attributes are measured for saplings (1.0-4.9 inches DBH). Each subplot is surrounded by a macroplot, on which rare events such as large trees are optionally measured. In addition to tree attributes, data are collected that describe the area where trees are located on each subplot and macroplot. If area attributes (e.g., ownership group, forest type, stocking) vary substantially across a plot, the land area within the plot is divided into distinct domains referred to as condition classes so that tree data can be properly associated with area classifications (Figure \ref{fig:plotDesign}). 

Plot-level estimates of forest attributes (e.g., tree biomass, forested area) are obtained by summing observations or estimates of the attribute, which are within the population of interest (e.g., live trees, private land), across the plot. Observations are multiplied by 1 if the attribute is within the population of interest, and 0 otherwise. Thus if all observations on a plot fall outside the population of interest, observations will sum to 0 for the plot. For ratio estimates (e.g., tree biomass per unit area), it is possible to specify different populations of interest for the numerator and denominator attributes.

The FIA program uses an annual panel system to estimate both current inventory and change. Individual panels are represented by a subset of ground plots that are measured in the same year and represent complete spatial coverage across the population of interest \citep{McRoberts2016}. In the eastern US, inventory cycles consist of 5 or 7 annual panels, with 20 or 15 percent of ground plots measured in each year, respectively. In the western US, inventory cycles include 10 annual panels and thus 10 percent of ground plots are measured each year \citep{bechtold2005enhanced}. Variance reduction can often be achieved by combining current panels with data from previous panels, although FIA does not prescribe a core procedure for panel combination because a single approach is unlikely to be suitable for all estimation objectives or across a wide variety of spatial, temporal, and population conditions \citep{bechtold2005enhanced}. 

In \texttt{rFIA}, users may choose from one of four unique estimators that combine data from multiple panels or choose to return estimates from individual annual panels (ANNUAL), thereby forgoing panel combination entirely. The \enquote{temporally-indifferent} (TI) estimator is used by default, essentially pooling all panels within an inventory cycle into a large periodic inventory. The TI estimator is commonly used by other FIA estimation tools, like EVALIDator \citep{toolsData, popsUserGuide}. Alternatively, individual panels may be weighted by employing a moving-average estimator, including the simple moving average (SMA), linear moving average (LMA), or exponential moving average (EMA). The SMA applies equal weight to all annual panels (Eq. \ref{eq:sma}), while the LMA and EMA apply weights that decay linearly or exponentially as a function of time since measurement, respectively (Eq. \ref{eq:lma}-\ref{eq:ema}). In each case of the moving average, panel weights sum to one across an inventory cycle. Panel weights are computed as follows:

\begin{align}
  w_{p_,SMA}=\frac{1}{N} \label{eq:sma} \\
  w_{p_,LMA}=\frac{p}{\sum_{i=1}^{N} p_i} \label{eq:lma} \\
  w_{p_,EMA}=\frac{\lambda(1-\lambda)^{1-p}}{\sum_{i=1}^{N} \lambda(1-\lambda)^{1-p_i}} \label{eq:ema}
\end{align}

\noindent where $w_p$ is the constant positive weight for annual panel, $p$, described as the sequential index of the panel within an inventory cycle (i.e., $p=1,2,\ldots,N$), $N$ is the total number of annual panels in the inventory cycle, and $\lambda$ is a decay parameter, ranging between zero and one, that controls the behavior of the exponential weighting function (Eq. 3). As $\lambda$ approaches 1, panel weights become nearly evenly distributed across the inventory cycle, and estimates produced by the EMA will approach those of the SMA. As $\lambda$ approaches 0, weights are skewed towards the most recent panel and estimates of the EMA will approach those of individual annual panels. The inclusion of $\lambda$ makes the EMA the most versatile estimator offered in \texttt{rFIA}.

In general, sample variance is minimized when panel weights are evenly distributed across the inventory cycle (e.g., SMA), with the trade-off of introducing temporal lag-bias by weighting recent measurements  the same as older measurements. Hence, equal weighting schemes could be undesirable in settings where variable values change over time. Alternatively, applying lower weights to less recent panels reduces the effective sample size of the inventory, effectively increasing variance but reducing temporal smoothing by favoring more recent measurements. We advise users of \texttt{rFIA} to be aware of this trade-off between precision and temporal smoothing when considering various estimators offered in the package.

\subsection{Software testing}
We have conducted extensive validation for all estimated attributes for small and large areas against EVALIDator \citep{toolsData}. Here we present an abbreviated version of a validation for the state of Connecticut in the year 2018 using the \enquote{temporally-indifferent} estimator (Table \ref{tab:evalidator}). Current status estimates were produced for live trees with DBH $\geq$ 1.0 inch and annual change estimates were produced for all stems with DBH $\geq$ 5.0 inches. Down woody material estimates were produced for the 1000 hour fuel class (coarse woody debris). Total forest area estimates were produced using plots containing a forested condition. All estimates and associated sampling errors produced by \texttt{rFIA} match those produced by EVALIDator. A detailed description of the validation described in Table \ref{tab:evalidator}, including code used to produce estimates using \texttt{rFIA}, can be found in  Appendix A.

\begin{table}[t]
  \begin{center}
    \caption{Comparison of estimates produced by rFIA and EVALIDator for select forest attributes. Estimates from both tools were produced using the \enquote{temporally-indifferent} estimator.}
    \label{tab:evalidator}
    \begin{tabular}{l S[table-format=3.2] S[table-format=3.2] S[table-format=3.2] S[table-format=3.2]}
    \hline
    Forest Attribute & \multicolumn{2}{c}{rFIA} & \multicolumn{2}{c}{EVALIDator} \\

     & \multicolumn{1}{c}{\textit{Estimate}} & \multicolumn{1}{c}{\textit{SE}} & \multicolumn{1}{c}{\textit{Estimate}} & \multicolumn{1}{c}{\textit{SE}} \\
     \hline
       Live tree abundance ($trees/ acre$) & 432.63 & 4.46 & 432.63 & 4.46  \\
       Live tree basal area ($ft^2/ acre$) & 121.19 & 2.13 & 121.19 & 2.13 \\
       Live tree merchantable volume ($ft^3/ acre$) & 2625.99 & 2.65 & 2625.99 & 2.65 \\
       Live tree sawlog volume ($ft^3/ acre$) & 1648.77 & 3.56 & 1648.77 & 3.56 \\
       Live tree aboveground biomass ($tons/ acre$) & 75.99 & 2.38 & 75.99 & 2.38 \\
       Live tree aboveground carbon ($tons/ acre$) & 37.99 & 2.38 & 37.99 & 2.38 \\
       Annual net biomass growth ($tons/ acre/ year$)* & 1.06 & 6.39 & 1.06 & 6.39 \\
       Annual mortality ($trees/ acre/ year$)* & 1.47 & 6.93 & 1.47 & 6.93 \\
       Annual removals ($trees/ acre/ year$)* & 0.36 & 31.09 & 0.36 & 31.09 \\
       Coarse woody material volume ($ft^3/ acre$) & 299.87 & 23.92 & 299.87 & 23.92 \\
       Coarse woody material biomass ($tons/ acre$) & 3.08 & 25.25 & 3.08 & 25.25 \\
       Coarse woody material carbon ($tons/ acre$) & 1.52 & 25.14 & 1.52 & 25.14\\
       Total forest area ($acres \: x \: 10^{-3}$) & 1789.61 & 2.29 & 1789.61 & 2.29 \\
      \hline

       \hline
    \end{tabular}
  \end{center}
\end{table}

\section{\texttt{rFIA} package features}

\begin{figure}[t]
    \centering
    \caption{Schematic diagram of the features of \texttt{rFIA}: the blue boxes represent decision points for the user; green boxes represent data or results that are produced and/or processed by \texttt{rFIA} functions, which are represented in the orange boxes.}
    \label{fig:useDiagram}
    \includegraphics[width=8cm]{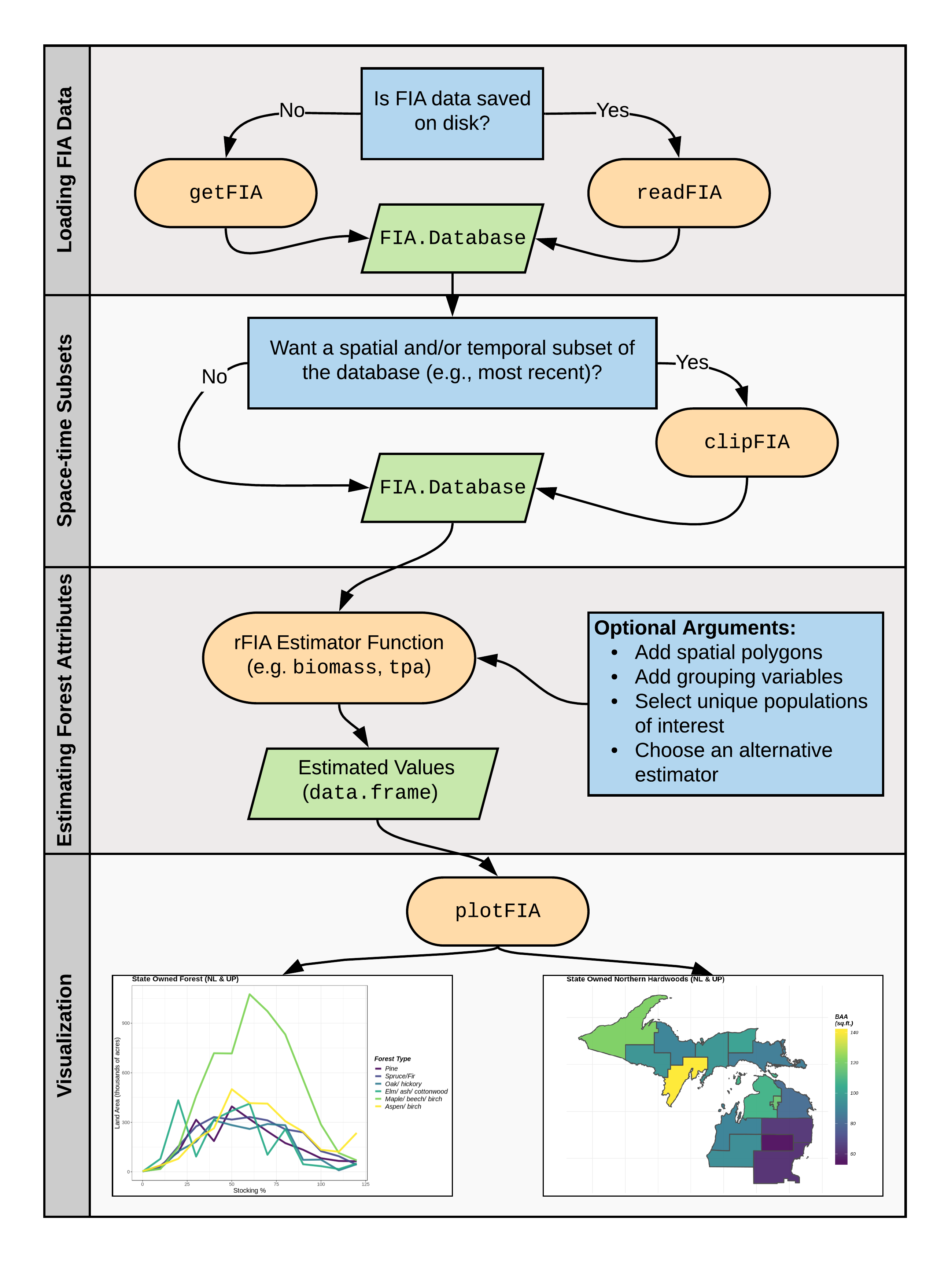}
\end{figure}

\texttt{rFIA} is capable of estimating more forest attributes from FIA data than any other publicly available tool and offers unmatched flexibility in defining unique populations of interest and producing space-time indexed summaries of forest attributes. Users can install the released version of \texttt{rFIA} from CRAN (v0.1.0, 28 October 2019), or alternatively install the development version from Github (https://github.com/hunter-stanke/rFIA) using \texttt{devtools} \citep{devtools2019}. Table \ref{tab:coreFunctions} depicts a list of core functions available in \texttt{rFIA}. A schematic diagram for using \texttt{rFIA} to produce population estimates of forest attributes can be found in Figure \ref{fig:useDiagram}.

\begin{table}[t]
  \begin{center}
    \caption{List of core functions available within \texttt{rFIA}.}
    \label{tab:coreFunctions}
    \begin{tabular}{l l} 
    \hline
      Function & Description \\
      \hline
       \texttt{area} & Estimate land area in various classes \\
       \texttt{biomass} & Estimate volume, biomass, and carbon stocks of standing trees \\
       \texttt{clipFIA} & Spatial and temporal queries for FIA data\\
       \texttt{diversity} & Estimate species diversity\\
       \texttt{dwm} & Estimate volume, biomass, and carbon stocks of down woody material\\
       \texttt{findEVALID} & Lookup Evaluation IDs (EVALIDs) by year and evaluation types\\
       \texttt{getFIA} & Download FIA data, load into R, and optionally save to disk\\
       \texttt{growMort} & Estimate recruitment, mortality, and harvest rates\\
       \texttt{invasive} & Estimate areal coverage of invasive species\\
       \texttt{plotFIA} & Produce static and animated plots of FIA summaries\\
       \texttt{readFIA} & Load FIA database into R environment from disk\\
       \texttt{seedling} & Estimate abundance of seedlings\\
       \texttt{standStruct} & Estimate forest structural stage distributions\\
       \texttt{tpa} & Estimate abundance of standing trees\\
       \texttt{vitalRates} & Estimate live tree growth rates\\
       \texttt{writeFIA} & Write in-memory FIA database to disk\\
       \hline
    \end{tabular}
  \end{center}
\end{table}

\subsection{Loading FIA data}
Users can automatically download state subsets of the FIA database \citep{toolsData}, load these data into R, and optionally save to disk using \texttt{getFIA}. Simply specify the state abbreviation code for the state(s) of interest as a character vector in the \texttt{states} argument of \texttt{getFIA} (e.g., \texttt{states = "CT"} for Connecticut). Alternatively, \texttt{getFIA} can be used to download, load, and optionally save select table(s) contained in the FIA database by specifying the names of the tables of interest in the \texttt{tables} argument (e.g., \texttt{tables = c("TREE", "PLOT")} for the \texttt{TREE} and \texttt{PLOT} tables). All FIA data are downloaded from the FIA DataMart \citep{toolsData} and saved as comma-delimited text files in a local directory. 

If FIA data are already saved on disk as comma-delimited text files, \texttt{readFIA} can be used to load these data into R. \texttt{readFIA} requires that FIA data files maintain original FIA naming conventions (as downloaded from the FIA DataMart) for referential integrity. All FIA data files should be stored in a single directory with no sub-directories for proper loading with \texttt{readFIA}. We recommend using \texttt{getFIA} to download and save new FIA data and \texttt{readFIA} to reload these data in future R sessions. All data loaded with \texttt{getFIA} or \texttt{readFIA} are stored as a modified list object called a \texttt{FIA.Database}. All common list operations are valid for the \texttt{FIA.Database} object class, and individual FIA tables can be accessed using the \texttt{\$} operator.

\subsection{Spatial and temporal subsets}
Spatial and/or temporal subsets of a \texttt{FIA.Database} may be implemented with \texttt{clipFIA} if users are interested in producing estimates for select areal regions and/or time periods contained within the spatial-temporal extent of the \texttt{FIA.Database}. Such subsets are not required to use \texttt{rFIA} estimator functions (listed in Table \ref{tab:estFunctions}) to produce estimates of forest attributes. However, limiting the spatial and temporal extent of the query region will conserve memory and decrease processing time. 

Users may subset a \texttt{FIA.Database} to the boundaries of a spatial polygon object by specifying the name of the spatial object in the \texttt{mask} argument of \texttt{clipFIA}. All spatial polygon classes from the \texttt{sp} and \texttt{sf} packages are supported. To obtain the most recent subset of a \texttt{FIA.Database} by reporting year, specify \texttt{mostRecent = TRUE} in \texttt{clipFIA} (\texttt{TRUE} by default). If a \texttt{FIA.Database} contains multiple states with different reporting schedules, setting \texttt{mostRecent = TRUE} will return the data necessary to produce estimates for the most recent reporting year in each state. Alternatively, users may specify \texttt{matchEval = TRUE} to obtain a subset of data associated with reporting years which are common among all states in the \texttt{FIA.Database}.

\subsection{Estimating forest attributes}
\texttt{rFIA} includes a range of estimator functions designed to produce population and plot-level estimates directly from \texttt{FIA.Database} objects. All estimator functions share a similar design, although minor nuances exist between functions due to variation in the forms of estimates they invoke (Table \ref{tab:estFunctions}). We will demonstrate the functionality of \texttt{rFIA} estimator functions using \texttt{biomass}, a function designed to estimate volume, biomass, and carbon of standing trees. 

\begin{table}[t]
  \begin{center}
    \caption{List of core arguments available in rFIA estimator functions.}
    \label{tab:estFunctions}
    \begin{tabular}{l c c c c c c} 
    \hline
       & \texttt{bySpecies} & \texttt{bySizeClass} & \texttt{treeDomain} & \texttt{areaDomain} & \texttt{tidy} & \texttt{method}\\
      \hline
       \texttt{area} & & & & * & & *  \\
       \texttt{biomass} & * & * & * & * & &*  \\
       \texttt{diversity} & 	& * & * & * & & *  \\
       \texttt{dwm} &  &  &  & * & * & * \\
       \texttt{growMort}  & * & * & * & * & & *\\
       \texttt{invasive} & (default) & & & * & & *\\
       \texttt{seedling} &  * &  & * & * & & * \\
       \texttt{standStruct} &  &  &  & * & * & *\\
       \texttt{tpa} &  * & * & * & * & & * \\
       \texttt{vitalRates} & * & * & * & * & & *\\
       \hline
    \end{tabular}
  \end{center}
\end{table}

To produce population estimates for the region contained within the spatial extent of a \texttt{FIA.Database}, specify the name of the \texttt{FIA.Database} as the \texttt{db} argument of any \texttt{rFIA} estimator function. All \texttt{rFIA} estimator functions return population estimates and associated sampling errors for each reporting year in the \texttt{FIA.Database} by default. To return estimates at the plot-level, specify \texttt{byPlot = TRUE}. In this case, estimates will be returned for each occasion that an individual plot was measured.

\begin{knitrout}
\definecolor{shadecolor}{rgb}{0.969, 0.969, 0.969}\color{fgcolor}\begin{kframe}
\begin{alltt}
\hlcom{## Load state subset for Rhode Island,}
\hlcom{## included with rFIA}
\hlkwd{data}\hlstd{(}\hlstr{"fiaRI"}\hlstd{)}
\hlcom{## Estimates for entire state}
\hlkwd{biomass}\hlstd{(}\hlkwc{db} \hlstd{= fiaRI)}
\end{alltt}
\begin{verbatim}
# A tibble: 5 x 19
   YEAR NETVOL_ACRE SAWVOL_ACRE BIO_AG_ACRE BIO_BG_ACRE BIO_ACRE
  <int>       <dbl>       <dbl>       <dbl>       <dbl>    <dbl>
1  2014       2396.       1356.        68.0        13.5     81.6
2  2015       2438.       1385.        69.1        13.7     82.9
3  2016       2491.       1419.        70.6        14.0     84.6
4  2017       2500.       1422.        70.8        14.1     84.8
5  2018       2491.       1419.        70.4        14.0     84.4
# ... with 13 more variables: CARB_AG_ACRE <dbl>, CARB_BG_ACRE <dbl>,
#   CARB_ACRE <dbl>, NETVOL_ACRE_SE <dbl>, SAWVOL_ACRE_SE <dbl>,
#   BIO_AG_ACRE_SE <dbl>, BIO_BG_ACRE_SE <dbl>, BIO_ACRE_SE <dbl>,
#   CARB_AG_ACRE_SE <dbl>, CARB_BG_ACRE_SE <dbl>, CARB_ACRE_SE <dbl>,
#   nPlots_VOL <dbl>, nPlots_AREA <dbl>
\end{verbatim}
\begin{alltt}
\hlcom{## Plot-level estimates}
\hlkwd{biomass}\hlstd{(}\hlkwc{db} \hlstd{= fiaRI,} \hlkwc{byPlot} \hlstd{=} \hlnum{TRUE}\hlstd{)}
\end{alltt}
\begin{verbatim}
# A tibble: 352 x 11
   PLT_CN  YEAR NETVOL_ACRE SAWVOL_ACRE BIO_AG_ACRE BIO_BG_ACRE BIO_ACRE
    <dbl> <int>       <dbl>       <dbl>       <dbl>       <dbl>    <dbl>
1 1.45e14  2009          0           0          0           0        0  
2 1.45e14  2009          0           0          0           0        0  
3 1.45e14  2009          0           0          0           0        0  
4 1.45e14  2009       2335.       1481.        70.4        13.6     84.0
5 1.45e14  2009       2760.       1690.        86.3        16.7    103. 
# ... with 347 more rows, and 4 more variables: CARB_AG_ACRE <dbl>,
#   CARB_BG_ACRE <dbl>, CARB_ACRE <dbl>, nStems <int>
\end{verbatim}
\end{kframe}
\end{knitrout}


\subsubsection{Grouped Estimates}
Often it is useful to produce estimates grouped by discrete categories, such as species, forest type, ownership group, or diameter class. \texttt{rFIA} estimator functions can produce estimates grouped by fields contained in the \texttt{PLOT}, \texttt{COND}, and \texttt{TREE} tables of a \texttt{FIA.Database}. To produce grouped estimates, specify the name(s) of fields representing the grouping variable(s) as the \texttt{grpBy} argument of any estimator function:
\begin{knitrout}
\definecolor{shadecolor}{rgb}{0.969, 0.969, 0.969}\color{fgcolor}\begin{kframe}
\begin{alltt}
\hlcom{## Grouping by Forest Type}
\hlkwd{biomass}\hlstd{(}\hlkwc{db} \hlstd{= fiaRI,} \hlkwc{grpBy} \hlstd{= FORTYPCD)}
\end{alltt}
\begin{verbatim}
# A tibble: 120 x 20
   YEAR FORTYPCD NETVOL_ACRE SAWVOL_ACRE BIO_AG_ACRE BIO_BG_ACRE BIO_ACRE
  <int>    <int>       <dbl>       <dbl>       <dbl>       <dbl>    <dbl>
1  2014      103       4030.       3362.        77.8        17.0     94.8
2  2014      104       3467.       2311.        74.3        16.1     90.4
3  2014      105       3876.       3403.        85.2        18.1    103. 
4  2014      167       2545.       1589.        60.6        13.2     73.8
# ... with 116 more rows, and 13 more variables: CARB_AG_ACRE <dbl>,
#   CARB_BG_ACRE <dbl>, CARB_ACRE <dbl>, NETVOL_ACRE_SE <dbl>,
#   SAWVOL_ACRE_SE <dbl>, BIO_AG_ACRE_SE <dbl>, BIO_BG_ACRE_SE <dbl>,
#   BIO_ACRE_SE <dbl>, CARB_AG_ACRE_SE <dbl>, CARB_BG_ACRE_SE <dbl>,
#   CARB_ACRE_SE <dbl>, nPlots_VOL <dbl>, nPlots_AREA <dbl>
\end{verbatim}
\end{kframe}
\end{knitrout}
If more than one grouping variable is provided to \texttt{grpBy}, grouping will occur hierarchically based on the order variable names are listed. In addition to \texttt{grpBy}, some \texttt{rFIA} estimator functions include \texttt{bySpecies} and \texttt{bySizeClass} arguments for convenience. Set either of these arguments as \texttt{TRUE} to produced estimates grouped by species and/or 2.0 inch DBH classes, respectively. Alternatively, users may specify \texttt{grpBy = SPCD} to group estiamtes by species and use the \texttt{makeClasses} function to define their own diameter classes. More information on variable definitions in the FIA Database can be found in the FIA Database Description and User Guide for Phase 2 \citep{burrill2018forest}. 

\subsubsection{Unique populations of interest}
In many cases, the population of interest is a subset of that represented by the full \texttt{FIA.Database}. For example, we may be interested in producing estimates for live stems greater than 12.0 inches DBH on state-owned land. Within \texttt{biomass} (and most other estimator functions), we can use the \texttt{treeDomain} and \texttt{areaDomain} arguments to describe our population of interest in terms of variables contained in the \texttt{PLOT}, \texttt{COND}, and \texttt{TREE} tables of a \texttt{FIA.Database}:
\begin{knitrout}
\definecolor{shadecolor}{rgb}{0.969, 0.969, 0.969}\color{fgcolor}\begin{kframe}
\begin{alltt}
\hlcom{## Live tree biomass (DBH > 12") on state land}
\hlkwd{biomass}\hlstd{(fiaRI,} \hlkwc{treeDomain} \hlstd{= DIA} \hlopt{>} \hlnum{12}\hlstd{,} \hlkwc{areaDomain} \hlstd{=} \hlstd{OWNCD} \hlstd{==} \hlnum{31}\hlstd{)}
\end{alltt}
\begin{verbatim}
# A tibble: 5 x 19
   YEAR NETVOL_ACRE SAWVOL_ACRE BIO_AG_ACRE BIO_BG_ACRE BIO_ACRE
  <int>       <dbl>       <dbl>       <dbl>       <dbl>    <dbl>
1  2014       1368.       1095.        34.8        6.96     41.7
2  2015       1396.       1094.        35.6        7.13     42.7
3  2016       1487.       1145.        38.0        7.61     45.6
4  2017       1430.       1101.        37.0        7.39     44.4
5  2018       1441.       1111.        37.3        7.45     44.7
# ... with 13 more variables: CARB_AG_ACRE <dbl>,
#   CARB_BG_ACRE <dbl>, CARB_ACRE <dbl>, NETVOL_ACRE_SE <dbl>,
#   SAWVOL_ACRE_SE <dbl>, BIO_AG_ACRE_SE <dbl>,
#   BIO_BG_ACRE_SE <dbl>, BIO_ACRE_SE <dbl>, CARB_AG_ACRE_SE <dbl>,
#   CARB_BG_ACRE_SE <dbl>, CARB_ACRE_SE <dbl>, nPlots_TREE <dbl>,
#   nPlots_AREA <dbl>
\end{verbatim}
\end{kframe}
\end{knitrout}
Here \texttt{treeDomain} describes the population of interest for the numerator (tree biomass) and \texttt{areaDomain} the population of interest for the denominator (state-owned forest land area). In each case, the argument takes the form of a logical predicate that is defined in terms of the variables in \texttt{PLOT}, \texttt{TREE}, and/or \texttt{COND} tables of the \texttt{FIA.Database}. Multiple conditions can be combined within either argument using \texttt{\&} or \texttt{|} symbols (and/or, respectively). 

\subsubsection{Grouping by user-defined areal units}
To produce estimates grouped by unique areal units, specify the name of a spatial polygon object (class \texttt{sp} or \texttt{sf}) defining the areal units of interest as the \texttt{polys} argument of any \texttt{rFIA} estimator function. All FIA data are automatically re-projected to match the projection of the input spatial polygon object prior to initiating estimation procedures. All fields originally contained in the input spatial polygon object will be preserved in the estimates output by the estimator function. To return estimates as an \texttt{sf} spatial polygon object, specify \texttt{returnSpatial = TRUE}. 

\subsubsection{Alternative estimators}
All \texttt{rFIA} estimator functions use the \enquote{temporally-indifferent} estimator by default (\texttt{method = "TI"}) for consistency with other FIA estimation tools, like EVALIDator \citep{toolsData, popsUserGuide}. As an alternative, users may set the \texttt{method} argument to \texttt{"SMA"}, \texttt{"LMA"}, \texttt{"EMA"}, or \texttt{"ANNUAL"}, to use the simple moving average, linear moving average, exponential moving average, or annual estimator, respectively. If using the exponential moving average, users may also modify the exponential decay parameter, $\lambda$ (Eq. \ref{eq:sma}), with the \texttt{lambda} argument. By default, \texttt{lambda} is set to \texttt{0.5}, although can be set to any value on the interval $(0,1)$. If multiple values are specified as the \texttt{lambda} argument, one unique set of estimates will be returned for each unique value of \texttt{lambda}. For example, \texttt{lambda = seq(from=0.1, to=0.9, by=0.1)} will produce nine unique sets of estimates, grouped by \texttt{lambda}:
\begin{knitrout}
\definecolor{shadecolor}{rgb}{0.969, 0.969, 0.969}\color{fgcolor}\begin{kframe}
\begin{alltt}
\hlcom{## Most recent year in Rhode Island}
\hlcom{## Multiple lambda values}
\hlkwd{biomass}\hlstd{(}\hlkwd{clipFIA}\hlstd{(fiaRI),} \hlkwc{method} \hlstd{= "EMA"} \hlstd{,}
        \hlkwc{lambda} \hlstd{= seq(from=0.1, to=0.9, by=0.1)} 
\end{alltt}
\begin{verbatim}
# A tibble: 9 x 20
  lambda  YEAR NETVOL_ACRE SAWVOL_ACRE BIO_AG_ACRE BIO_BG_ACRE
   <dbl> <int>       <dbl>       <dbl>       <dbl>       <dbl>
1    0.1  2018       2451.       1387.        68.8        13.7
2    0.2  2018       2455.       1397.        69.2        13.7
3    0.3  2018       2449.       1400.        69.2        13.7
4    0.4  2018       2432.       1393.        68.9        13.7
5    0.5  2018       2405.       1376.        68.4        13.6
# ... with 4 more rows, and 13 more variables: BIO_ACRE <dbl>, 
#   CARB_AG_ACRE <dbl>, CARB_BG_ACRE <dbl>, CARB_ACRE <dbl>, 
#   NETVOL_ACRE_SE <dbl>, SAWVOL_ACRE_SE <dbl>, BIO_AG_ACRE_SE <dbl>, 
#   BIO_BG_ACRE_SE <dbl>, BIO_ACRE_SE <dbl>, CARB_AG_ACRE_SE <dbl>,
#   CARB_BG_ACRE_SE <dbl>, CARB_ACRE_SE <dbl>, nPlots_TREE <dbl>, 
#   nPlots_AREA <dbl>
\end{verbatim}
\end{kframe}
\end{knitrout}

\subsubsection{Parallelization}
Parallel processing is available in all \texttt{rFIA} estimator functions using the \texttt{nCores} argument. \texttt{nCores} indicates the number of physical cores to be used and may be set to any positive integer up to the number of physical cores available on a given machine. Serial processing is implemented by default (\texttt{nCores = 1}). Parallelization may substantially reduce memory during processing. Thus, users should consider implementing serial processing if computational resources are limited. If implementing parallel processing, we recommend users set \texttt{nCores} to one less than the number of physical cores available on their machine to ensure computational resources are available for other processes (e.g., OS, browsers).

\section{Case study - Michigan ash decline}
We demonstrate the utility of \texttt{rFIA} for forest resource monitoring by assessing the decline of ash (\emph{Fraxinus} spp.) populations across the Lower Peninsula of Michigan following the establishment of emerald ash borer (\emph{Agrilus planipennis Fairmaire}). Emerald ash borer is an invasive forest insect that was first discovered in southeastern Michigan in 2002. The insect spread quickly across the state and is considered one of the most destructive and costly forest insects to invade the United States \citep{poland2006emerald, aukema2011economic}. Using the \enquote{temporally indifferent} estimator in \texttt{rFIA}, we estimated annual changes in live ash trees per acre (TPA) and tree mortality rates (mortality TPA per year) by county during the interval 2006-2018. To highlight differences among alternative design-based estimators implemented in \texttt{rFIA}, we use each estimator to assess changes in ash sawlog stocks (board feet per acre) across the entire Lower Peninsula from 2000 to 2018 and compare their relative performance (\texttt{lambda = 0.5} used for EMA estimator). All estimates were produced for white ash (\emph{Fraxinus americana L.}), green ash (\emph{Fraxinus pennsylvanica Marshall}), and black ash (\emph{Fraxinus nigra Marshall}) trees $\geq$ 5 inches DBH. 

\begin{figure}[t]
    \centering
    \caption{Changes in live ash TPA (left) and annual mortality rates (right), with associated sampling errors (bottom; 67\% confidence), across counties in the Lower Peninsula of Michigan following establishment of emerald ash borer. Missing values are shaded in gray. All plots were produced using the \texttt{plotFIA} function. }
    \label{fig:mi_ash}
    \includegraphics[width=8cm]{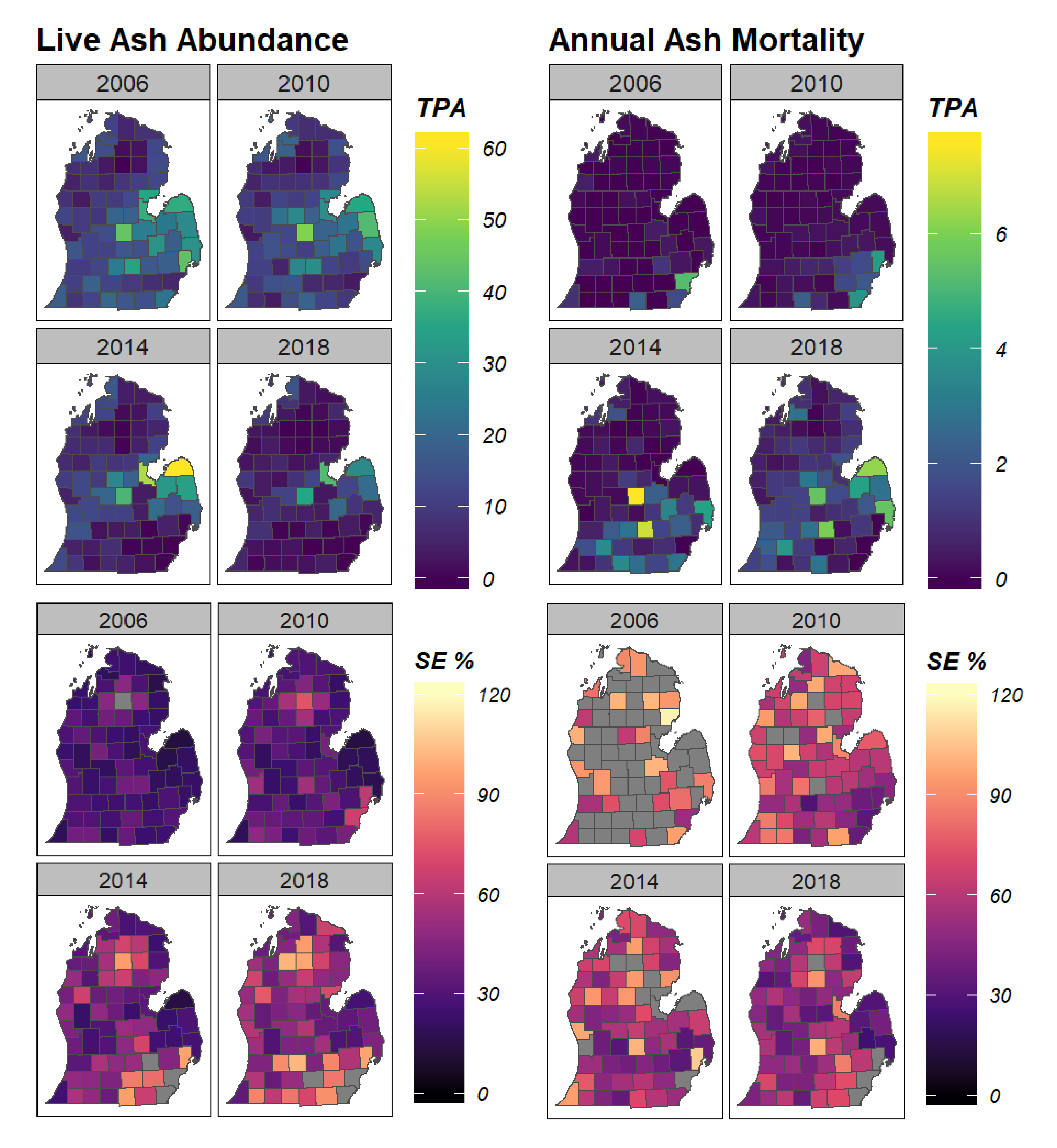}
\end{figure}

\begin{figure}[t]
    \centering
    \caption{Changes in ash sawlog volume (top; board feet (BF) per acre) and associated sampling errors (bottom; 67\% confidence) for each design-based estimator, across the Lower Peninsula of Michigan from 2000 to 2018. All plots were produced using the \texttt{plotFIA} function. }
    \label{fig:ash_estimators}
    \includegraphics[width=8cm]{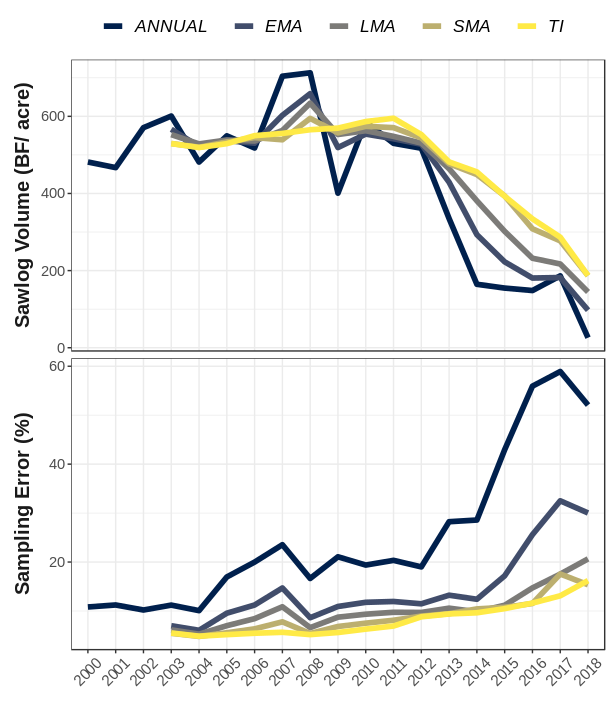}
\end{figure}

A rapid decline of ash populations across Michigan counties is evident from our analysis. Live ash TPA decreased by 61\%  and annual mortality increased by 583\% across the study region from 2006 to 2018 (\enquote{temporally-indifferent} estimator). Elevated mortality rates and population decline are apparent in the southeastern portion of the state soon after the establishment of emerald ash borer in the region (Figure \ref{fig:mi_ash}, 2006 and 2010). In years following, ash population decline and elevated mortality became evident across the Lower Peninsula, likely associated with the rapid expansion of emerald ash borer across the state (Figure \ref{fig:mi_ash}, 2014 and 2018). 

Sampling error is directly related to the number of non-zero observations (i.e., FIA ground plots) used to compute estimates of each variable, and cannot be determined if less than two non-zero observations are available. Decreasing precision (i.e., increased sampling errors) associated with live TPA (Figure \ref{fig:mi_ash}; bottom-left) and sawlog volume (Figure \ref{fig:ash_estimators}; bottom) estimates indicates that the number of live ash trees located on FIA ground plots decreased over time. Ash mortality appeared to be a relatively rare event prior to the expansion of emerald ash borer across Michigan, as many counties recorded fewer than two observations of ash mortality per sampling period (gray shaded counties in bottom right of Figure \ref{fig:mi_ash}). The increased frequency of ash mortality observations across Michigan counties, and associated reduction in sampling error, provides support for the trend of increased mortality rates that is evident following the establishment of emerald ash borer across the state.

A trade-off between temporal smoothing and precision is evident in comparing the behavior of design-based estimators implemented in \texttt{rFIA} (Figure \ref{fig:ash_estimators}). For example, the annual estimator (ANNUAL) appears to indicate a sharper and earlier decline (lower temporal smoothing) in the ash sawlog resource than the \enquote{temporally-indifferent} (TI) and simple moving average (SMA) estimators, though the annual estimator consistently produces higher sampling errors (lower precision) (Figure \ref{fig:ash_estimators}). Estimators that give more weight to recent observations, like the annual (100\% of weight on most recent observation), linear moving average (LMA), and exponential moving average (EMA) tend to exhibit lower temporal lag-bias but at the cost of higher variance. In contrast, estimators that distribute weight more evenly across observations with respect to time, like the \enquote{temporally-indifferent} (no annual weighting, treated as periodic inventory) and simple moving average tend to exhibit higher temporal smoothing but lower variance. In practice, we recommend users carefully consider this trade-off and choose an estimator that achieves both their desired precision standards and an acceptably low level of temporal lag-bias.

\section{Extensions/limitations}
In the future, we hope to expand \texttt{rFIA} to include model-based and model-assisted techniques, allowing users to leverage various forms of auxiliary data to improve estimation of forest attributes. Specifically, we hope to improve estimation within small domains (e.g., spatial/ temporal extents, rare forest attributes) by implementing a number of indirect and composite small area estimators. Further, we hope to improve the value of the FIA Database for detection and analysis of rapid changes in forest health through the implementation of Kalman filters as time-series estimators. 

Program R is an \enquote{in-memory} application by design, and \texttt{rFIA} requires all input data to be held in RAM. Hence, users may experience memory management challenges when computational resources are limited. Memory constraints may inhibit some users from loading or processing large subsets of FIA data (relative to available computational resources), and high memory usage may result in decreased computational efficiency. Future development seeks to leverage \texttt{dplyr} \citep{wickham2015dplyr} back ends for external database engines such as SQLite and Apache Spark, thus reducing memory limitations and potentially offering substantial improvements in computational efficiency for large data. 

Exact coordinates of FIA ground plots are not available in the public version of the FIA Database to protect privacy rights of private land owners and preserve the ecological integrity of ground plots \citep{tinkham2018applications, mcroberts2005enhanced}. Plot locations are randomly displaced up to 1 km from their true locations (i.e., \enquote{fuzzed}) and the coordinates of up to 20\% of plots on private land in each county are exchanged (i.e., \enquote{swapped}) \citep{bechtold2005enhanced, tinkham2018applications}. The effects of using \enquote{fuzzed and swapped} plot coordinates are thought to be negligible for design-based estimation across large regions (800,000 acres and larger) \citep{mcroberts2005fuzzswap}, however uncertainty in true plot locations may produce results within unknown amounts of error when used in conjunction with other spatially explicit data layers for model-based estimation \citep{sabor2007adding}. Hence, we encourage \texttt{rFIA} users to consider this uncertainty when using design-based estimation for small populations, and when producing plot-level estimates to be used with spatially explicit auxiliary data in a model-based or model-assisted estimation framework.

\section{Conclusion}
The FIA database is among the most valuable ecological datasets in the world, though its complexity limits access for many potential users. We developed \texttt{rFIA} to simplify the estimation of forest attributes using FIA data, intending to provide a flexible, yet powerful toolset which is accessible to all R users. Ultimately, we hope that \texttt{rFIA} will improve the accessibility and relevance of the FIA database and promote use of FIA data among a larger, more diverse audience. We encourage users to apply \texttt{rFIA} widely and report any issues and/or desired extensions on our active issues page (https://github.com/hunter-stanke/rFIA/issues).

\section{Acknowledgements}
This work was supported by National Science foundation grants DMS-1916395, EF-1241874, EF-1253225, National Aeronautics and Space Administration Carbon Monitoring System program, and United States National Park Service and Forest Service.

\clearpage


\bibliography{lit.bib}



\clearpage
\appendix
\section{}

This appendix is intended to serve as further reference to section 2.3, Software Testing. We have conducted extensive validation of \texttt{rFIA} for all available attributes against EVALIDator. Here we present the code used to produce an abbreviated version of a validation for the state of Connecticut in the year 2018 (Table 1, section 2.3). 

First we download the state subset of the FIA Database for Connecticut and then subset the database to only include observations necessary to compute estimates for the year 2018:

\begin{knitrout}
\definecolor{shadecolor}{rgb}{0.969, 0.969, 0.969}\color{fgcolor}\begin{kframe}
\begin{alltt}
\hlcom{## Load rFIA}
\hlkwd{library}\hlstd{(rFIA)}

\hlcom{## Download data}
\hlstd{ct} \hlkwb{<-} \hlkwd{getFIA}\hlstd{(}\hlstr{'CT'}\hlstd{)}

\hlcom{## 2018 Subset}
\hlstd{ct} \hlkwb{<-} \hlkwd{clipFIA}\hlstd{(ct,} \hlkwc{mostRecent} \hlstd{=} \hlnum{FALSE}\hlstd{,} \hlkwc{evalid} \hlstd{=} \hlkwd{findEVALID}\hlstd{(ct,} \hlkwc{year} \hlstd{=} \hlnum{2018}\hlstd{))}
\end{alltt}
\end{kframe}
\end{knitrout}

\subsection{Live tree abundance (trees/acre)}
\emph{EVALIDator}
\begin{itemize}[noitemsep]
  \item Land Basis: Forestland
  \item Numerator: Tree Number
  \item Denominator: Area
  \item Estimate: Number of live trees (at least 1 inch d.b.h./d.r.c.), in trees, on forest land
  \item Evaluation: 092018N CONNECTICUT 2012;2013;2014;2015;2016;2017;2018
  \item Row Variable: EVALID
  \item Column Variable: EVALID
\end{itemize}

\emph{rFIA}
\begin{knitrout}
\definecolor{shadecolor}{rgb}{0.969, 0.969, 0.969}\color{fgcolor}\begin{kframe}
\begin{alltt}
\hlcom{# Trees per acre}
\hlstd{tpa_ct} \hlkwb{<-} \hlkwd{tpa}\hlstd{(ct)}
\hlstd{tpa_ct}\hlopt{$}\hlstd{TPA}
\end{alltt}
\end{kframe}
\end{knitrout}

\subsection{Live tree basal area (sq.ft./acre)}
\emph{EVALIDator}
\begin{itemize}[noitemsep]
  \item Land Basis: Forestland
  \item Numerator: Tree basal area
  \item Denominator: Area
  \item Estimate: Basal area of live trees (at least 1 inch d.b.h./d.r.c.), in square feet, on forest land
  \item Evaluation: 092018N CONNECTICUT 2012;2013;2014;2015;2016;2017;2018
  \item Row Variable: EVALID
  \item Column Variable: EVALID
\end{itemize}

\emph{rFIA}
\begin{knitrout}
\definecolor{shadecolor}{rgb}{0.969, 0.969, 0.969}\color{fgcolor}\begin{kframe}
\begin{alltt}
\hlcom{# Basal area per acre (BAA)}
\hlstd{tpa_ct} \hlkwb{<-} \hlkwd{tpa}\hlstd{(ct)}
\hlstd{tpa_ct}\hlopt{$}\hlstd{BAA}
\end{alltt}
\end{kframe}
\end{knitrout}

\subsection{Live tree merchantable volume (cu.ft./acre)}
\emph{EVALIDator}
\begin{itemize}[noitemsep]
  \item Land Basis: Forestland
  \item Numerator: Tree volume
  \item Denominator: Area
  \item Estimate: Net merchantable bole volume of live trees (at least 5 inch d.b.h./d.r.c.), in cubic feet, on forest land
  \item Evaluation: 092018N CONNECTICUT 2012;2013;2014;2015;2016;2017;2018
  \item Row Variable: EVALID
  \item Column Variable: EVALID
\end{itemize}

\emph{rFIA}
\begin{knitrout}
\definecolor{shadecolor}{rgb}{0.969, 0.969, 0.969}\color{fgcolor}\begin{kframe}
\begin{alltt}
\hlcom{# Net volume per acre}
\hlstd{bio_ct} \hlkwb{<-} \hlkwd{biomass}\hlstd{(ct)}
\hlstd{bio_ct}\hlopt{$}\hlstd{NETVOL_ACRE}
\end{alltt}
\end{kframe}
\end{knitrout}

\subsection{Live tree sawlog volume (cu.ft./acre)}
\emph{EVALIDator}
\begin{itemize}[noitemsep]
  \item Land Basis: Forestland
  \item Numerator: Tree volume
  \item Denominator: Area
  \item Estimate: Net sawlog volume of live trees, in cubic feet, on forest land
  \item Evaluation: 092018N CONNECTICUT 2012;2013;2014;2015;2016;2017;2018
  \item Row Variable: EVALID
  \item Column Variable: EVALID
\end{itemize}

\emph{rFIA}
\begin{knitrout}
\definecolor{shadecolor}{rgb}{0.969, 0.969, 0.969}\color{fgcolor}\begin{kframe}
\begin{alltt}
\hlcom{# Sawlog volume per acre}
\hlstd{bio_ct} \hlkwb{<-} \hlkwd{biomass}\hlstd{(ct)}
\hlstd{bio_ct}\hlopt{$}\hlstd{SAWVOL_ACRE}
\end{alltt}
\end{kframe}
\end{knitrout}

\subsection{Live tree aboveground biomass (tons/acre)}
\emph{EVALIDator}
\begin{itemize}[noitemsep]
  \item Land Basis: Forestland
  \item Numerator: Tree dry weight
  \item Denominator: Area
  \item Estimate: Aboveground biomass of live trees (at least 1 inch d.b.h./d.r.c.), in short tons, on forest land
  \item Evaluation: 092018N CONNECTICUT 2012;2013;2014;2015;2016;2017;2018
  \item Row Variable: EVALID
  \item Column Variable: EVALID
\end{itemize}

\emph{rFIA}
\begin{knitrout}
\definecolor{shadecolor}{rgb}{0.969, 0.969, 0.969}\color{fgcolor}\begin{kframe}
\begin{alltt}
\hlcom{# Sawlog volume per acre}
\hlstd{bio_ct} \hlkwb{<-} \hlkwd{biomass}\hlstd{(ct)}
\hlstd{bio_ct}\hlopt{$}\hlstd{BIO_AG_ACRE}
\end{alltt}
\end{kframe}
\end{knitrout}

\subsection{Live tree aboveground carbon (tons/acre)}
\emph{EVALIDator}
\begin{itemize}[noitemsep]
  \item Land Basis: Forestland
  \item Numerator: Tree carbon
  \item Denominator: Area
  \item Estimate: Aboveground carbon of live trees (at least 1 inch d.b.h./d.r.c.), in short tons, on forest land
  \item Evaluation: 092018N CONNECTICUT 2012;2013;2014;2015;2016;2017;2018
  \item Row Variable: EVALID
  \item Column Variable: EVALID
\end{itemize}

\emph{rFIA}
\begin{knitrout}
\definecolor{shadecolor}{rgb}{0.969, 0.969, 0.969}\color{fgcolor}\begin{kframe}
\begin{alltt}
\hlcom{# Sawlog volume per acre}
\hlstd{bio_ct} \hlkwb{<-} \hlkwd{biomass}\hlstd{(ct)}
\hlstd{bio_ct}\hlopt{$}\hlstd{CARB_AG_ACRE}
\end{alltt}
\end{kframe}
\end{knitrout}

\subsection{Annual net biomass growth (tons/acre/year)}
\emph{EVALIDator}
\begin{itemize}[noitemsep]
  \item Land Basis: Forestland
  \item Numerator: Annual net growth dry weight
  \item Denominator: Area
  \item Estimate: Average annual net growth of aboveground biomass of trees (at least 5 inch d.b.h./d.r.c.), in short tons, on forest land
  \item Evaluation: 092018Y CONNECTICUT 2012;2013;2014;2015;2016;2017;2018
  \item Row Variable: EVALID
  \item Column Variable: EVALID
\end{itemize}

\emph{rFIA}
\begin{knitrout}
\definecolor{shadecolor}{rgb}{0.969, 0.969, 0.969}\color{fgcolor}\begin{kframe}
\begin{alltt}
\hlcom{# Annual biomass growth per acre}
\hlstd{vr_ct} \hlkwb{<-} \hlkwd{vitalRates}\hlstd{(ct)}
\hlstd{vr_ct}\hlopt{$}\hlstd{BIO_GROW_AC}
\end{alltt}
\end{kframe}
\end{knitrout}

\subsection{Annual mortality (trees/acre/year)}
\emph{EVALIDator}
\begin{itemize}[noitemsep]
  \item Land Basis: Forestland
  \item Numerator: Annual mortality number
  \item Denominator: Area
  \item Estimate: Average annual mortality of trees (at least 5 inch d.b.h./d.r.c.), in trees, on forest land
  \item Evaluation: 092018Y CONNECTICUT 2012;2013;2014;2015;2016;2017;2018
  \item Row Variable: EVALID
  \item Column Variable: EVALID
\end{itemize}

\emph{rFIA}
\begin{knitrout}
\definecolor{shadecolor}{rgb}{0.969, 0.969, 0.969}\color{fgcolor}\begin{kframe}
\begin{alltt}
\hlcom{# Annual mortality TPA}
\hlstd{gm_ct} \hlkwb{<-} \hlkwd{growMort}\hlstd{(ct)}
\hlstd{gm_ct}\hlopt{$}\hlstd{MORT_TPA}
\end{alltt}
\end{kframe}
\end{knitrout}

\subsection{Annual removals (trees/acre/year)}
\emph{EVALIDator}
\begin{itemize}[noitemsep]
  \item Land Basis: Forestland
  \item Numerator: Annual removal number
  \item Denominator: Area
  \item Estimate: Average annual removals of trees (at least 5 inch d.b.h./d.r.c.), in trees, on forest land
  \item Evaluation: 092018Y CONNECTICUT 2012;2013;2014;2015;2016;2017;2018
  \item Row Variable: EVALID
  \item Column Variable: EVALID
\end{itemize}

\emph{rFIA}
\begin{knitrout}
\definecolor{shadecolor}{rgb}{0.969, 0.969, 0.969}\color{fgcolor}\begin{kframe}
\begin{alltt}
\hlcom{# Annual removals TPA}
\hlstd{gm_ct} \hlkwb{<-} \hlkwd{growMort}\hlstd{(ct)}
\hlstd{gm_ct}\hlopt{$}\hlstd{REMV_TPA}
\end{alltt}
\end{kframe}
\end{knitrout}

\subsection{Coarse woody debris volume (cu.ft./acre)}
\emph{EVALIDator}
\begin{itemize}[noitemsep]
  \item Land Basis: Forestland
  \item Numerator: Down woody material volume
  \item Denominator: Area
  \item Estimate: Total volume of CWD, in cubic feet, on forest land
  \item Evaluation: 092018N CONNECTICUT 2012;2013;2014;2015;2016;2017;2018
  \item Row Variable: EVALID
  \item Column Variable: EVALID
\end{itemize}

\emph{rFIA}
\begin{knitrout}
\definecolor{shadecolor}{rgb}{0.969, 0.969, 0.969}\color{fgcolor}\begin{kframe}
\begin{alltt}
\hlcom{# CWD volume}
\hlstd{dwm_ct} \hlkwb{<-} \hlkwd{dwm}\hlstd{(ct)}
\hlstd{dwm_ct}\hlopt{$}\hlstd{VOL_ACRE[dwm_ct}\hlopt{$}\hlstd{FUEL_TYPE} \hlopt{==} \hlstr{'1000HR'}\hlstd{]}
\end{alltt}
\end{kframe}
\end{knitrout}

\subsection{Coarse woody debris biomass (tons/acre)}
\emph{EVALIDator}
\begin{itemize}[noitemsep]
  \item Land Basis: Forestland
  \item Numerator: Down woody material biomass
  \item Denominator: Area
  \item Estimate: Weight of CWD, in short tons, on forest land
  \item Evaluation: 092018N CONNECTICUT 2012;2013;2014;2015;2016;2017;2018
  \item Row Variable: EVALID
  \item Column Variable: EVALID
\end{itemize}

\emph{rFIA}
\begin{knitrout}
\definecolor{shadecolor}{rgb}{0.969, 0.969, 0.969}\color{fgcolor}\begin{kframe}
\begin{alltt}
\hlcom{# CWD biomass}
\hlstd{dwm_ct} \hlkwb{<-} \hlkwd{dwm}\hlstd{(ct)}
\hlstd{dwm_ct}\hlopt{$}\hlstd{BIO_ACRE[dwm_ct}\hlopt{$}\hlstd{FUEL_TYPE} \hlopt{==} \hlstr{'1000HR'}\hlstd{]}
\end{alltt}
\end{kframe}
\end{knitrout}

\subsection{Coarse woody debris carbon (tons/acre)}
\emph{EVALIDator}
\begin{itemize}[noitemsep]
  \item Land Basis: Forestland
  \item Numerator: Down woody material carbon
  \item Denominator: Area
  \item Estimate: Carbon of CWD, in short tons, on forest land
  \item Evaluation: 092018N CONNECTICUT 2012;2013;2014;2015;2016;2017;2018
  \item Row Variable: EVALID
  \item Column Variable: EVALID
\end{itemize}

\emph{rFIA}
\begin{knitrout}
\definecolor{shadecolor}{rgb}{0.969, 0.969, 0.969}\color{fgcolor}\begin{kframe}
\begin{alltt}
\hlcom{# CWD carbon}
\hlstd{dwm_ct} \hlkwb{<-} \hlkwd{dwm}\hlstd{(ct)}
\hlstd{dwm_ct}\hlopt{$}\hlstd{CARB_ACRE[dwm_ct}\hlopt{$}\hlstd{FUEL_TYPE} \hlopt{==} \hlstr{'1000HR'}\hlstd{]}
\end{alltt}
\end{kframe}
\end{knitrout}

\subsection{Total forest area ($acres \: x \: 10^{-3}$)}
\emph{EVALIDator}
\begin{itemize}[noitemsep]
  \item Land Basis: Forestland
  \item Numerator: Area
  \item Denominator: None
  \item Estimate: Area in forestland, in acres
  \item Evaluation: 092018N CONNECTICUT 2012;2013;2014;2015;2016;2017;2018
  \item Row Variable: EVALID
  \item Column Variable: EVALID
\end{itemize}

\emph{rFIA}
\begin{knitrout}
\definecolor{shadecolor}{rgb}{0.969, 0.969, 0.969}\color{fgcolor}\begin{kframe}
\begin{alltt}
\hlcom{# Total Forest Area}
\hlstd{fa_ct} \hlkwb{<-} \hlkwd{area}\hlstd{(ct)}
\hlstd{fa_ct}\hlopt{$}\hlstd{AREA_TOTAL}
\end{alltt}
\end{kframe}
\end{knitrout}

\end{document}